\documentclass[twocolumn,superscriptaddress,floatfix,prl,showpacs]{revtex4}
\usepackage[dvips]{graphicx}

\newcommand{\oph}{\omega_{\mbox{\scriptsize ph}}}
\newcommand{\PRL}{Phys. Rev. Lett. }
\newcommand{\PRB}{Phys. Rev. B }

\begin{document}
\title{Localization-delocalization transition of a  
polaron near an impurity}
\author{A.~S.~Mishchenko}
\affiliation{Cross-Correlated Materials Research Group (CMRG), ASI,
RIKEN, Wako 351-0198, Japan}
\affiliation{RRC ``Kurchatov Institute'', 123182, Moscow, Russia}

\author{N.~Nagaosa}
\affiliation{Cross-Correlated Materials Research Group (CMRG), ASI,
RIKEN, Wako 351-0198, Japan}
\affiliation{Department of Applied Physics, The University of Tokyo,
7-3-1 Hongo, Bunkyo-ku, Tokyo 113, Japan} 
 
\author{A.~Alvermann}
\affiliation{Institut f{\"u}r Physik,
Ernst-Moritz-Arndt-Universit{\"a}t Greifswald, 17489 Greifswald,
Germany}

\author{H.~Fehske}
\affiliation{Institut f{\"u}r Physik,
Ernst-Moritz-Arndt-Universit{\"a}t Greifswald, 17489 Greifswald,
Germany}

\author{G.~De~Filippis}
\affiliation{Coherentia-CNR-INFM and Dip. di Scienze Fisiche - 
Universit\`{a} di Napoli Federico II - I-80126 Napoli, Italy}

\author{V.~Cataudella}
\affiliation{Coherentia-CNR-INFM and Dip. di Scienze Fisiche - 
Universit\`{a} di Napoli Federico II - I-80126 Napoli, Italy}

\author{O.~P.~Sushkov}
\affiliation{School of Physics, University of New South Wales, 
Sydney 2052, Australia}

\begin{abstract}
{We solve the problem of polaron localization on an attractive impurity
by means of direct-space Diagrammatic Monte Carlo implemented 
for the system in the thermodynamic limit. In particular we determine the 
ground state phase diagram in dependence on the electron-phonon coupling and 
impurity potential strength for the whole  phonon frequency range.  
Including the quantum phonon dynamics we find and characterize a new phase  
which is missing in the zero phonon-frequency limit (adiabatic approximation), 
where self-trapped polarons are not localized at shallow impurities.  
We predict and show that in the vicinity of the localization transition 
a region with a mixture of weak- and strong-coupling spectral response is 
realized.      
}
\end{abstract}


\maketitle

A general approach to the theoretical description of a particle in a bulk 
medium coupled both to bosonic excitations and the potential of imperfections 
is an important but notoriously hard problem that poses a real challenge even 
to modern nonperturbative approaches \cite{Olle03}. 
As yet only approximate results, 
relying, e.g., on dynamical-mean field theory exist. 
A central question in this context is the formation of three-dimensional 
(3D) polarons at impurities, or the Anderson localization of polarons in 
disordered media~\cite{DISPOL,Bro,AlKo}.
The overall importance of the physics of electron-phonon interaction in 
doped materials makes this issue of general interest for different areas of 
physics and technology. As a matter of fact 
the interplay between disorder and interaction effects is an important issue 
for contemporary materials design. 
For example high temperature superconductors \cite{Lee06,Gun08,Vit07_OCtJ08}
or materials with colossal magnetoresistance \cite{CMR} are doped Mott 
insulators where besides the coupling to bosonic excitations (phonons and 
magnons) disorder is present. 

In this Letter we present the exact solution to the polaron problem in the 
presence of an attractive impurity in a 3D material.
The accepted model for that situation
is given by the Hamiltonian $H=H^{(0)}+ H^{(1)}$ with
\begin{eqnarray}
H^{(0)}&=& 
- U c_{\bf 0}^{\dagger} c_{\bf 0} +
\omega_{\mbox{\scriptsize ph}} \sum_{\bf i} b_{\bf i}^{\dagger} b_{\bf i} \; , 
\label{h0}\\
H^{(1)}&= &
-t\sum_{\langle {\bf i},{\bf j} \rangle} c_{\bf i}^{\dagger} c_{\bf j} 
- \gamma \sum_i (b_{\bf i}^{\dagger}+b_{\bf i})  
c_{\bf i}^{\dagger} c_{\bf i} 
  \; .
\label{int}
\end{eqnarray}
In $H^{(0)}$, $U$ is the attractive impurity potential for the 
electron $c^\dagger_0$ at  site $0$ and  
$b^\dagger_i$ creates a dispersionless optical phonon with  frequency 
$\omega_{\mbox{\scriptsize ph}}$ at Wannier $i$. $H^{(1)}$ describes the 
electron transfer $\propto t$ between nearest neighbor sites 
 and local Holstein coupling 
to the phonons  $\propto \gamma$.

In the absence of electron-phonon (el-ph) coupling ($\gamma=0$),
the critical $U$  for particle localization at the impurity is 
$U_c(\gamma=0) \approx 3.96$ \cite{KostSlet}; all energies are 
measured in units of $t$ hereafter.
In the adiabatic approximation (AA), setting 
$\omega_{\mbox{\scriptsize ph}} = 0$, the phase diagram in $U-\lambda$
coordinates, with the dimensionless coupling constant 
$\lambda=\gamma^2 / (6t\omega_{\mbox{\scriptsize ph}})$, was established  
in Ref.~\cite{ShiToy79}.
The phase boundary in AA, separating delocalized polaron states from localized 
ones, crosses the $U$-axis at 
$U_c(\gamma=0)$ and the $\lambda$-axis at 
$\lambda_c(U_c=0,\omega_{\mbox{\scriptsize ph}}=0) \approx 0.9$.
The latter crossing is a confusing property of the AA phase diagram 
since it implies that for el-ph couplings  
$\lambda > \lambda_c(U_c=0,\oph=0)$ the polaron is localized 
even when $U=0$. 
Quite the contrary, a particle is never localized in a translationally 
invariant 
lattice ($U=0$) with quantum phonons ($\omega_{\mbox{\scriptsize ph}}>0$).
Instead the particle undergoes only a crossover from the weak-coupling light 
polaron to a strong-coupling heavy polaron with small radius around a 
self-trapping coupling $\lambda_{\mbox{\scriptsize ST}}$ 
\cite{MPS_Rashba_Pekar}. 
The AA erroneously equates $\lambda_{\mbox{\scriptsize ST}}$  with the 
critical el-ph coupling strength $\lambda_c$ required for polaron 
localization at $U \ne 0$. 
Therefore drastic differences between the exact result and that 
obtained in AA are expected, especially at small $U_c$. 

Having this delicate situation in mind, we decided to study the full 
Hamiltonian (\ref{h0})-(\ref{int}) with quantum phonons. To this end 
we employ a new scheme combining the Diagrammatic 
Monte Carlo (DMC) method  in direct space \cite{MPSS} and the Stochastic 
Optimization Method for analytic continuation \cite{MPSS,Olle}
which provides the approximation-free solution of the above problem 
without finite size errors and in zero temperature limit.  
Calculating the charge density distribution (CDD) around the impurity and 
the local density of states (LDOS) on the impurity site we 
establish the exact localization phase diagram for different 
phonon frequencies. 
We characterize two novel polaronic regimes in a system with 
impurities. 
The polaron at small $U$ can be self-trapped though extended and not yet
confined by shallow impurities. 
Another regime, arising near the critical parameters for localization
at the impurity, shows spectroscopic response like a mixture
of spectra typical for weak, intermediate, and strong coupling.    
 
The direct space DMC method~\cite{Macridin2004}
can provide the direct space Green 
functions (GFs) in imaginary time ($\tau$)
representation at zero temperature
$G_{\bf{ij}}(\tau)=\langle \mbox{vac} |
c_{\bf j}(\tau) c_{\bf i}^{\dagger} | \mbox{vac} \rangle$ 
for the Hamiltonian (\ref{h0}-\ref{int}) 
by Feynman diagram expansion in the 
interaction representation 
\begin{equation} 
G_{\bf{ij}}(\tau)= 
\left\langle e^{- \tau H^{(0)}}
\widehat{T}_{\tau}  \left[ \langle c_{\bf j}(\tau) c_{\bf i}^{\dagger}
\exp \left\{
- \int_0^{\tau} \! \! H^{(1)}(\tau') d\tau' 
\right]\right\} 
\right\rangle .
\label{macrid}
\end{equation}
The implementation of DMC \cite{Macridin2004} requires to keep in computer 
memory all GFs $\{G_{\bf{ij}}(\tau) \}$ in  direct space, which restricts 
the lattice to about $25 \times 25$ sites.
To avoid this size limitation we calculate only quantities related  
to on-site GFs $G_{\bf{ii}}(\tau)$. 
With our implementation of DMC we are able to calculate 
the on-site GFs at zero temperature for a $10^8 \times 10^8 \times 10^8$ 
lattice, thereby avoiding any finite-size or finite temperature errors.
A slight modification of 
Eq.~(\ref{macrid}),
\begin{equation} 
n({\bf i})= 
\left\langle \frac{e^{- \beta H^{(0)}}}{Z}
\widehat{T}_{\tau}  \left[ \langle c_{\bf i}(\beta) c_{\bf i}^{\dagger}
\exp \left\{
- \int_0^{\beta} \! \! H^{(1)}(\tau') d\tau' 
\right]\right\} 
\right\rangle ,
\label{dens}
\end{equation}
introduces the estimator for the CDD at temperature 
$T=\beta^{-1}$. 
To make a calculation of the CDD feasible, we 
collect its statistics in a cube with $40^3$ number of sites. 
Note that this strategy does not introduce finite-size errors because 
only the $\tau=0,\beta$ points of the partition function loop are confined to 
the $40^3$-cube while the diagrams are free to sample all $(10^8)^3$ sites. 

The CDD estimator is effective for locating the localization parameters for 
large $U$ only but, because of the requirement of finite temperatures, 
fails at small $U \ll t$. 
Note that the path-integral quantum Monte-Carlo algorithm \cite{Kornilo}, 
which is another method relevant for the problem formulated above \cite{AlKo}, 
has serious precision limits for the same reason.
Hence, the only rigorous method to locate the localization point
in the infinite system is to calculate the on-site zero temperature GF 
$G_{\bf{ii}}(\tau)$, determine the LDOS 
$L_{\bf i}(\omega) = - \pi^{-1} \mathrm{Im} G_{\bf{ii}}(\omega)$ by 
analytic continuation \cite{MPSS,Olle}, and check for the presence of a 
bound state in the LDOS $L_{\bf 0}(\omega)$ at the impurity site. 

To validate the new implementation of the DMC technique, we located the 
critical $U_c(\lambda=0) \approx 3.96$ by calculating the CDD, 
normalized to unity at the impurity site, around the impurity. 
It occurred that for $U \le U_c$ the charge density does not decrease 
exponentially with distance from the impurity while for $U> U_c$ it does. 
Perfect agreement is found between CDD obtained by DMC and that 
obtained in Ref.~\cite{KostSlet}. 
For $U$ close to $U_c$, however, determination of the LDOS $L_{\bf 0}(\omega)$ 
is a much more precise method, since the CDD requires finite temperatures.

\begin{figure}[thb]
\begin{center}
\includegraphics[width=8cm]{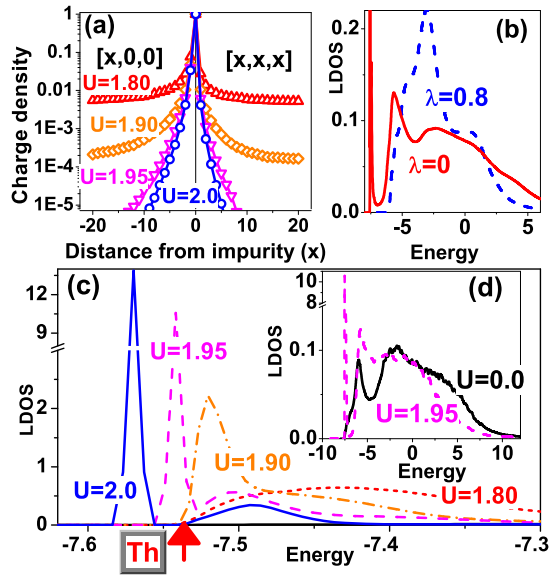}
\end{center}
\caption{\label{fig:fig1} (color online) 
Charge density at $T=0.001$ (a) and LDOS at the impurity site
for $T=0$ (c,d) for different values of $U$ at $\oph=2$ and 
$\lambda=0.8$.
The arrow in panel (c) indicates the lower
threshold (Th) of the spectrum at $\lambda=0.8$ and $U=0$.
Panel (b) shows LDOS at the impurity site at $U=2$, $\oph=2$ for 
$\lambda=0$ (dashed line) and $\lambda=0.8$ (solid line).      
Statistical errorbars in panel (a) are less than $10^{-5}$ (i.e. much 
less than the point size).
} 
\end{figure}

\begin{figure}[bth]
\begin{center}
\includegraphics[width=8cm]{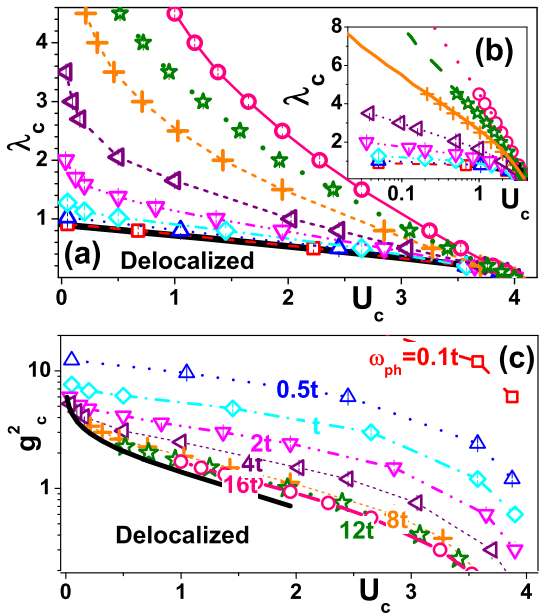}
\end{center}
\caption{\label{fig:fig2} (color online)  Phase boundaries between delocalized 
(left lower corner) and localized states of a polaron in 
$\lambda_c - U_c$ (a,b) and $g^2_c - U_c$ (c) coordinates:
adiabatic limit $\omega_{\mbox{\scriptsize ph}}=0$ 
(thick solid line in panel (a)), 
$\omega_{\mbox{\scriptsize ph}}=0.1$ (squares),      
$0.5$ (triangles up),
$1$ (diamonds),
$2$ (triangles down),      
$4$ (triangles left),      
$8$ (crosses),      
$12$ (stars), 
$16$ (circles).     
The thick solid curve in panel (c) is obtained from Eq.~(\ref{ucg}).
Solid, dashed and dotted lines in panel (b) are results obtained by 
the CBS method \cite{noi} 
for $\omega_{\mbox{\scriptsize ph}}=8, 12$, and $16$.
The values of $\lambda_c$ are set exactly while the errorbars
of the quantity $U_c$ are less than $3.0 \times 10^{-2}$.
} 
\end{figure}

First let us demonstrate how trapped polaron states 
are determined using the LDOS.
From the commutator $[H,c^\dagger_i]$ we find that, independent of the el-ph 
coupling $\lambda$, the first moment $M_1$ of the LDOS 
$L_{\bf i}(\omega)$ obeys 
\begin{equation}
M_1 = 
\int_{-\infty}^{\infty} \! \!d \omega \; \omega L_{\bf i}(\omega) 
= - U \delta_{{\bf i},{\bf 0}} .  
\label{sum}
\end{equation}
In accordance with the sum rule (\ref{sum}), the LDOS at the impurity site 
shifts to lower energies with increasing $U$ (Fig.~\ref{fig:fig1}c). 
The second moment $M_2=\int_{-\infty}^{\infty} 
\! \!d \omega \; \omega^2 L_{\bf i}(\omega)$ increases with $\lambda$ 
(Fig.~\ref{fig:fig1}b) and the overall LDOS broadens.
Figure~\ref{fig:fig1} shows the CDD (a) and LDOS (c,d) for various 
values of $U$  at fixed $\lambda=0.8$ and 
$\oph=2$. 
Let us start the discussion with the case $U=0$, where no localization is 
expected. 
We determine the lower border $E_{\mbox{\scriptsize Th}}(\lambda)$ of the 
LDOS $L_{\bf 0}(\omega)$ for given $\lambda$ (arrow in 
Fig.~\ref{fig:fig1}(c)).
Increasing $U$ the LDOS changes but there is no 
spectral density below  $E_{\mbox{\scriptsize Th}}(\lambda)$ up
to $U \le 1.90$ (Fig.~\ref{fig:fig1}(c)). 
The CDD around the impurity, in accordance with the absence of a bound state
in LDOS, does not show exponential decrease too (Fig.~\ref{fig:fig1}(a)).
This gives another confirmation for the method employed here.
In order to search for the localization-delocalization transition
we proceed to larger values of $U$. For $U \ge 1.95$, the bound state 
appears below the threshold   
$E_{\mbox{\scriptsize Th}}(\lambda)$ (Fig.~\ref{fig:fig1}(c)),
and the CDD decays exponentially (Fig.~\ref{fig:fig1}(a)).
In this way we obtain one transition point in the phase diagram 
(Fig.~\ref{fig:fig2}(a)),
here $U_c(\oph=2,\lambda=0.8) = 1.925 \pm 0.025$. 
Recall that, although the LDOS approach needs a very precise determination  
(compare (Fig.~\ref{fig:fig1}(c) and (d)), it is applicable 
for any values of $\lambda$ and $U$. 
On the contrary, the CDD method is fast but, due to requirement of 
finite temperature, not reliable at $U \ll 1$.   

Next the phase diagram for polaron localization is 
presented in Fig.~\ref{fig:fig2}.
Using $\lambda_c-U_c$ coordinates (Fig.~\ref{fig:fig2}(a), (b)),
we see how our exact solution differs from the adiabatic result 
($\oph \to 0$, thick solid curve) for finite $\oph$.
In Fig.~\ref{fig:fig2}(c), with $g^2-U_c$ coordinates ($g=\gamma/\oph$),
we show the deviation from the limiting phase boundary at 
$\oph \to \infty$ (thick solid curve),
\begin{equation}
U_c(\oph=\infty) = U_c(\gamma=0) \exp(-g^2_c).
\label{ucg}
\end{equation}
This relation is obtained by Lang-Firsov transformation,
which renormalizes hopping, and accordingly the critical $U_c$, as 
$t \to t \exp(-g^2)$.
Note that a exponential relation between $U_c$ and $g_c^2$ 
(or $\lambda_c=g_c^2 \oph / 6t$) 
is a characteristic property of 
the small $U_c \ll 1$ and large $\lambda>1$ regime (Fig.~\ref{fig:fig2}(b)),
since for large $\oph$, 
\begin{equation}
\lambda_c(U_c,\oph) \simeq
(\oph/6t) \ln \left[ U_c(\gamma=0)/U_c \right] + \mbox{const} \;.
\label{border}
\end{equation} 
It is indeed seen in Fig.~\ref{fig:fig2}(b) that the slope of the phase 
boundary  increases with $\oph$.

\begin{figure}[bth]
\begin{center}
\includegraphics[width=8cm]{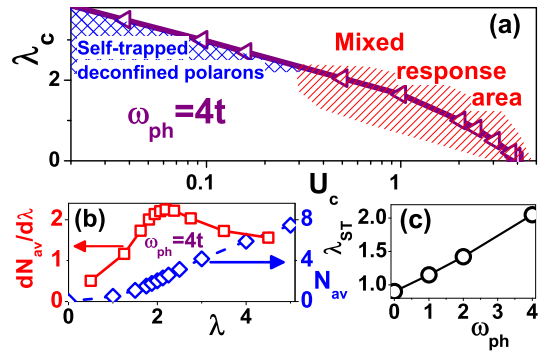}
\end{center}
\caption{\label{fig:fig3}  (color online) 
(a) Localization phase diagram and (b) average number of 
phonons $N_{\mbox{\scriptsize av}}$ and the derivative
$d N_{\mbox{\scriptsize av}} / d \lambda$   
for $\omega_{\mbox{\scriptsize ph}}=4$.
(c) Dependence of the self-trapping coupling $\lambda_{\mbox{\scriptsize ST}}$
on the phonon frequency $\omega_{\mbox{\scriptsize ph}}$. 
Errorbars are less than point size. 
} 
\end{figure}
 
We emphasize the excellent agreement of the novel DMC approach with 
the adiabatic limit \cite{ShiToy79} and antiadiabatic limit Eq.~(\ref{ucg})
results, as well as with the data obtained by the 
Coherent Basis States (CBS) method \cite{noi} for small $U$,
which proves the validity of the implementation. 
Note that the phase diagram obtained here is free from any substantial 
error and presents the first available solution of the Hamiltonian 
Eq.~(\ref{h0})-(\ref{int}) for all parameter regimes.

\begin{figure}[bth]
\begin{center}
\includegraphics[width=8cm]{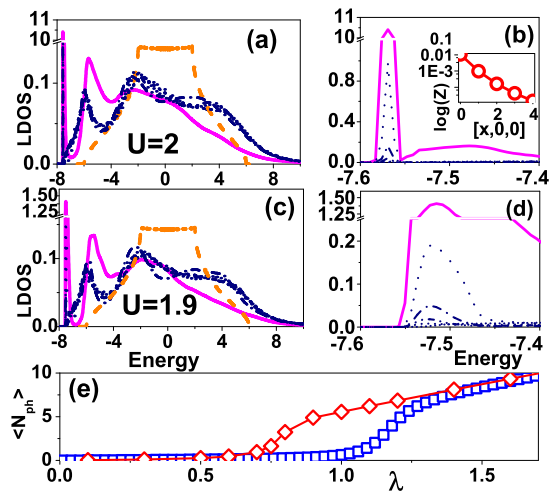}
\end{center}
\caption{\label{fig:fig4}  (color online) 
LDOS for a localized polaron (a,b) at $U=2$ and a delocalized polaron 
(c,d) at $U=1.9$, where $\lambda=0.8$ and $\omega_{\mbox{\scriptsize ph}}=2$.
Shown is the LDOS at the impurity site $(0,0,0)$ (solid line), nearest 
neighbor $(1,0,0)$ (dots), site $(2,0,0)$ (dash-dots), $(3,0,0)$  
(dash-dot-dots), $(4,0,0)$ (short-dashes), and for infinite distance 
from the impurity (short dots). 
The dashed line is the unperturbed LDOS ($\lambda=0$, $U=0$).
The inset in panel (b) gives the $Z$-factor of 
the LDOS $\delta$-peak of the bound state as a function of the distance 
to the impurity.
Panel (e) shows the average number of phonons at (diamonds) and far from 
the impurity (squares) for  $\oph=1$ and $U=1.4$. 
} 
\end{figure}

Let us finally discuss two essential features of the phase diagram,
which are entirely missing in the adiabatic approximation.
The first is realized at large el-ph couplings where the polarons are already 
self-trapped but not yet confined by shallow impurities
(cross-hatched region in Fig.~\ref{fig:fig3}(a)). 
The self-trapping coupling $\lambda_{\mbox{\scriptsize ST}}$, locating
the crossover from the weak- to strong-coupling regime, can be 
defined, e.g., as the maximum of the derivative of the average number of 
phonons in the ground state $N_{\mbox{\scriptsize av}}=
\left\langle b_{{\bf q}={\bf 0}}^{\dagger}  b_{{\bf q}={\bf 0}} \right\rangle$
with respect to the coupling constant $\lambda$ (Fig.~\ref{fig:fig3}(b,c)). 
For small enough $U \ll 1$, and any given $\oph$, one finds a 
sector in the phase diagram (Fig.~\ref{fig:fig3}(a)) where 
$\lambda_{\mbox{\scriptsize ST}}(\oph)<\lambda<\lambda_c(U,\oph)$.
This sector defines the phase of self-trapped deconfined polarons,
whose identification is obtained here for the first time.

The second novel feature appears at moderate values of $U$ close 
to the transition region between localized 
and extended states (line-shaded area in Fig.~\ref{fig:fig3}a). 
There, the spectral properties of a polaron are strongly position-dependent.  
In Fig.~\ref{fig:fig4} we show the LDOS calculated at, and in the vicinity of
the impurity.
Both for a localized (a,b) and extended 
(c,d) polaron the LDOS at the impurity site strongly 
differs from that at the nearest neighbor site in the low-energy region
(b,d). 
On the contrary, the overall features of LDOS at the nearest neighbor are very 
similar to those at infinite distance from the defect (Fig.~\ref{fig:fig4}a,c).
This property points out how strongly the spectral properties depend on the 
value of the impurity potential at a given lattice site. 
Comparison of the average number of phonons
at the impurity with that in infinite distance to the impurity 
(Fig.~\ref{fig:fig4}e) shows that for a wide range of parameters 
the lattice is weakly distorted far from 
the impurity while it is strongly deformed near the impurity.
This demonstrates how impurities enhance the formation of small polarons. 

As a consequence it is expected that in a material with imperfections a 
mixture of behavior typical for weak-coupling polarons (far away from an 
impurity) and strong-coupling polarons (close to, or at, the impurity) occurs.
Even though the impurity concentration can be small, the induced changes, e.g. 
in the spectral response, can be drastic. 
For example, for $U=1.90$ and $\lambda=0.8$ (Fig.~\ref{fig:fig1}a) the charge 
density on the impurity site is four orders of magnitude larger
than in the bulk of the system. 
Therefore, even a small impurity concentration $n_i \approx 10^{-3}$ suffices 
to entirely change the spectral properties. 
For example, since photoemission \cite{tJpho}
and optical conductivity \cite{Optics} spectra are a very different for 
weak and strong-coupling, one can expect very rich mixture of the spectral 
responses. 

In conclusion, introducing the direct space diagrammatic Monte Carlo in 
the thermodynamic limit we presented the exact 
phase diagram for localization of a polaron at an
attractive impurity for all coupling strengths, values of the impurity 
potential, and phonon frequencies ranging from the adiabatic to the 
antiadiabatic regime. Most notably we characterize a novel phase 
where heavy polarons are mobile in the presence
of shallow impurities and predict complex spectral properties of 
the systems close to the localization-delocalization 
transition. The present DMC method can be easily generalized to 
study more general situations, e.g. systems with long-range  
particle hopping, impurities with long-range attractive/repulsive 
potentials, or interfaces and layered structures, demonstrating the 
potential for future research.

A.S.M.\ is supported by RFBR 07-02-00067a. 
N.N. was partly supported by the
Grant-in-Aids from under the Grant No.\ 15104006, No.\ 16076205, 
No. 17105002, No.\ 19048015, and NAREGI Nanoscience Project from 
the Ministry of Education, Culture, Sports, Science, and Technology, Japan.
A.A. and H.F. acknowledge support by DFG SFB 652.

\end{document}